\documentstyle[prl,aps,epsf,multicol,graphicx]{revtex}
\begin{document}
\title{\bf The Nature of the Condensate in Mass Transport Models}
\author { Satya N. Majumdar $^1$, M.R. Evans $^2$ and R.K.P. Zia $^3$}
\address{
{\small $^1$Laboratoire de Physique Th\'eorique et Mod\`eles Statistiques,
        Universit\'e Paris-Sud. B\^at. 100. 91405 Orsay Cedex. France}\\
{\small $^2$ School of Physics, University of Edinburgh,
Mayfield Road, Edinburgh, EH9 3JZ, United Kingdom}\\
{\small $^3$ Department of Physics and
Center for Stochastic Processes in Science and Engineering,
Virginia Tech, Blacksburg, VA 24061-0435, USA}}

\date{\today}

\maketitle

\begin{abstract}
We study the phenomenon of real space condensation in the steady state of a class of one dimensional
mass transport models. We derive the criterion for the occurrence of a condensation transition
and analyse the precise nature of the shape and the size of the condensate in the condensed phase. 
We find two distinct condensate regimes: one where the
condensate is gaussian distributed and the particle number
fluctuations scale normally as $L^{1/2}$ where $L$ is the system size,
and a second regime where the particle number fluctuations become
anomalously large and the condensate peak is non-gaussian. We interpret
these results within the framework of sums of random variables.
\noindent


\end{abstract}

\begin{multicols}{2}
Condensation transitions are ubiquitous in nature. For systems in
thermal equilibrium, clustering is well understood in terms of the
competition between entropy and energy (typically associated with
attractive interactions). More exotic are condensations in systems
with {\em no} interactions, e.g., Bose-Einstein's free (quantum)
particles. Less understood are such transitions in {\em
non-equilibrium }systems, in some of which even the concept of energy
is dubious. For example, condensation is known to occur in many mass
transport models, defined only by a set of rules of evolution, with no
clear `attraction' between the masses
\cite{SHZ,MRE00,E96,BBJ,OEC,MKB}. The relevance of these models lies in
their applicability to a broad variety of phenomena, e.g., traffic
flow \cite{traffic}
force propagation through granular media \cite{CLMNW}, granular flow
\cite{granular} and network dynamics \cite{DM}. Correspondingly, the
condensation transition describes jamming in traffic \cite{E96},
bunching of buses \cite{OEC}, clogging in pipes \cite{OEC},
coalescence of shaken steel balls \cite{granular} and condensation of edges in networks
\cite{DM}.

How such transitions arise is especially intriguing for one
dimensional ($d=1$) systems with {\em local} dynamical rules. A well known
example is the Zero-Range Process (ZRP) \cite{MRE00,GSS,Godreche} in
which masses hop from site to (the next) site according to some transfer
rule. In the steady state, a finite fraction of the total mass `condenses'
onto a single site when $\rho $, the global mass density, is increased
beyond a certain critical value: $\rho _c$. The system goes from a fluid
phase, where the mass at each site hovers around $\rho $, to a condensed
phase, where a fluid of density $\rho _c$ co-exists with a condensate
containing all the `excess' mass.

Though condensation in these systems share interesting analogies\cite{E96,MKB} with
the traditional Bose-Einstein condensation, there are important
differences. For example, here condensation occurs in real space and
in all dimensions. Moreover these systems are {\em non-equilibrium}
in the sense that they are defined by the dynamics, generally lack
a Hamiltonian and the stationary state is not specified 
by the usual Gibbs-Boltzmann distribution. There are two major
problems that one faces in the analysis of condensation phenomenon
in these systems. First, the stationary state itself often is very
difficult to determine and secondly, even if it is known such as
in ZRP, the analysis of condensation has so far been possible
only within a grand canonical enemble (GCE) where one is
already in the thermodynamic ($L\to \infty$) limit. While
the GCE approach correctly predicts when a condensation transition
can happen and even the value of the critical density $\rho_c$, it
fails to provide much insight into the `condensed' phase
($\rho>\rho_c$) itself. For that one needs to work in a canonical
ensemble with the system size $L$ finite, which has not been
possible so far. In this Letter, we show that both of these 
problems can be overcome in a general class of mass transport
models recently introduced by us\cite{EMZ}. This allows
us to explore the condensed phase in detail revealing 
rather rich physical behaviors, in particular the existence
of two different types of condensates.      


Our model is defined as follows:
a mass $m_i$ resides at each site $i$
of a $d=1$ periodic lattice of size $L$. At each time step, a portion,
$\tilde{m}_i\le m_i$, chosen from a  distribution $\phi
(\tilde{m}|m)$, is chipped off to site $i+1$. The dynamics
conserves the total mass $ M=\sum_{i=1}^Lm_i=\rho L$. 
The model  is general enough to include many previously
studied models as special cases\cite{EMZ}. Choosing the
chipping kernel $\phi
(\tilde{m}|m)$ appropriately,  recovers ZRP, the Asymmetric Random
Average Process \cite{ARAP} and the chipping model of\cite{MKB}. 
Moreover the model
encompasses both discrete and continuous time dynamics and
discrete and continuous mass.  In particular, it was
shown that the stationary state 
has a  simple, factorised form provided 
the kernel is of the form
$\phi(\tilde{m}|m) \propto u(\tilde{m}) v(m-\tilde{m})$ \cite{EMZ},
where $u(z)$ and $v(z)$ are arbitrary non-negative functions.
Then the joint distribution of mass in the steady state is
given by
\begin{equation}
P(m_1,\cdots ,m_L)=\frac{\prod_{i=1}^Lf(m_i)}{Z(M,L)}\,\delta \left(
\sum_{j=1}^Lm_j-M\right)   \label{prodm1}
\end{equation}
where
  $f(m)=\int_0^{m} d{\tilde m} u(\tilde m) v(m-\tilde m)$ and the `canonical partition function' 
$Z(M,L)$ is just the normalization
\begin{equation}
Z(M,L)=\prod_{i=1}^L\int_0^\infty dm_i\,f(m_i)\delta \left(
\sum_{j=1}^Lm_j-M\right) \;.  \label{Zcan}
\end{equation}
Note that (\ref{prodm1}) is 
 a product of single-site weights $f(m_i)$ but
 the $\delta$-function in (\ref{prodm1})
implies a fixed total mass thus
inducing correlations between sites and in general the single-site
mass {\em probability distribution}  $p(m) \neq f(m)$.

The dynamics of the model specifies the functions $u(\tilde m)$ and
$v(m-\tilde m)$, which in turn specify the steady state uniquely in
terms of weight function $f(m)$. Having determined the steady state, one
next turns to the issue of condensation. In particular, we ask: (i) 
when does a condensation transition occur
(ii) if condensation occurs, what is the precise nature of the condensate?

The factorization property allows (i) to be addressed rather easily within a
GCE framework -- \`a la Bose Einstein. 
The approach implies taking the $L\to \infty$ limit and setting
the single-site
mass distribution function $p(m) = f(m) {\em e}^{-\mu m}$
where $\mu$ is the chemical potential and is chosen
to fix the density $\rho = \int dm p(m) m$.
Thus, condensation must occur
for $\rho>\rho_c = \int dm f(m) m$ which is the maximum
allowed value of $\rho$ within the GCE.
Based on previous works on the  ZRP related case \cite{BBJ,OEC}, 
it is easy to show that a condensation
transition occurs if the single site weights decay for large $m$ as
\begin{equation}
f(m)\simeq A\,m^{-\gamma }\quad {\rm with}\quad \gamma >2  \label{fm1}\;.
\end{equation}
A simple  example of a chipping kernel which gives
such weights is furnished by
$u(\tilde{m}) = \exp(-a \tilde{m})$ and
$v(m-\tilde{m})= (1+m-\tilde{m})^{-\gamma}$ which  yield
$f(m) \simeq m^{-\gamma}/a$ for large $m$.
In the following, we stay with the choice of $f(m)$ in (\ref{fm1})
and set, without loss of generality, $%
\int_0^\infty f(m)dm=1$.

The GCE analysis correctly predicts the criterion for condensation and even the
critical density $\rho_c$, but provides little insight 
into the condensed phase itself where $\rho >\rho _c$. In this work, we
are able to explore the condensed phase by staying
within the  canonical ensemble and analyzing the mass
distribution $p(m)\equiv \int 
dm_2....dm_LP(m,\cdots
,m_L)\delta \left( \sum_{j=2}^Lm_j+m-M\right)$
in a finite system of size $L$.
Using (\ref{Zcan}), we have
\begin{equation}
p(m)=f(m)\frac{Z(M-m,L-1)}{Z(M,L)}.  \label{pm1}
\end{equation}
The rest of the letter is devoted to the analysis of $p(m)$ in (\ref{pm1})
with $f(m)$ given by (\ref{fm1}). We have two parameters $\gamma$ and 
$\rho$. Our goal is to show how 
the condensation
is  manifested by  different behaviors of $p(m)$ in different regions of
the $(\rho-\gamma)$ plane giving rise to a rich phase diagram in Fig. 2.  

First, consider the Laplace transform of (\ref{Zcan}): 
\begin{equation}
\int_0^\infty Z(M,L)e^{-sM}dM=\left[ g(s)\right] ^L\;,  \label{lt1}
\end{equation}
where $g(s)=\int_0^\infty f(m)e^{-sm}dm$. The main challenge is to invert (%
\ref{lt1}) for a given $f(m)$ and exploit its behavior to analyse $p(m)$.
Before proceeding to the general case, let us present
a case in which both $Z(M,L)$ and $p(m)$ can be obtained in closed form. We
choose $f(m)=2{\rm e}^{-1/m}m^{-5/2}/\sqrt{\pi }$, for which $%
g(s)=(1+2s^{1/2}){\rm e}^{-2s^{1/2}}$ 
and our results below show that
$\rho _c= -g^{\prime }(0)=2$. Now, (\ref{lt1}) can be inverted to provide the
closed form\cite{tbp}: 
\begin{equation}
Z(M,L)=B_{M,L}\,\left[ H_L(r)-\sqrt{M}H_{L-1}(r)\right] ,  \label{esc1}
\end{equation}
where $B_{M,L}=LM^{-(L+3)/2}\,e^{-L^2/M}/\sqrt{\pi }$, $r=(2L+M)/{2\sqrt{M}}$
and $H_k(r)$ is the Hermite polynomial of degree $k$. Substituting $Z(M,L)$
in (\ref{pm1}), we obtain $p(m)$ explicitly and plot the case of $L=100$ in
Fig. 1. The transition from the subcritical ($\rho =1$), through the
critical ($\rho =2$), to the supercritical ($\rho =6$) cases, is clearly
seen: the condensate showing up as an additional, asymmetric 
bump for $\rho =6$. 
The explicit solution in this toy example provides us with useful insights
into what to expect in the general case.

\begin{figure}
\includegraphics[scale=0.4]{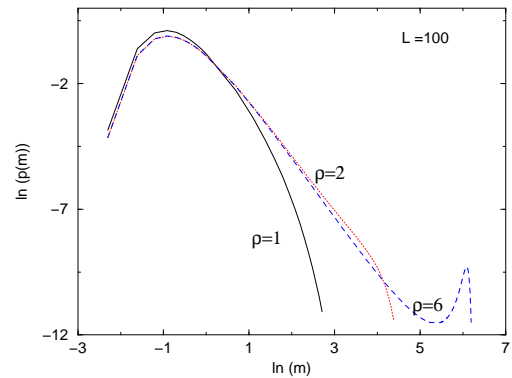}
\caption{The distribution $p(m)$ vs. $m$ for the exactly solvable
case, plotted using {\it Mathematica} for $L=100$ and $\rho=1$
(subcrtical), $\rho=\rho_c=2$ (critical) and $\rho=6$
(supercritical). The condensate shows up as an additional bump near
the tail of $p(m)$ in the supercritical case.}
\end{figure}

Before proceeding, let us summarise here our main results:
we refer to Fig. 1 for
typical forms of the mass distribution $p(m)$ and Fig. 2 for a
schematic phase diagram. In the subcritical regime the system is in a
fluid phase where the mass distribution decays exponentially
with decay length increasing with density.
At $\rho_c$ the distribution decays as a power law 
$p(m)\sim m^{-\gamma}$ and at $\rho >\rho_c$
the distribution develops an extra piece, representing the condensate,
centred around $M-L \rho_c$. By our analysis within the canonical
ensemble we show that this piece will have a `normal', gaussian form
when $\gamma>3$. When $\gamma<3$, however the condensate will have an
anomalous, asymmetric form as seen in Fig. 1 for $\rho =6$. In the following we
supply explicit expressions for these forms.

\begin{figure}
\includegraphics[scale=0.4]{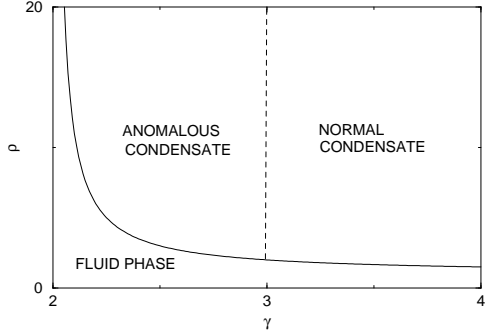}
\caption{Schematic phase diagram 
in the $\rho$--$\gamma$ plane. The full line represents
the critical density $\rho_c(\gamma)$}   
\end{figure}

We formally invert (\ref
{lt1}) using the Bromwich formula, 
\begin{equation}
Z(M,L)=\int_{s_0-i\infty }^{s_0+i\infty }\frac{ds}{2\pi i}\exp \left[
L\left( \ln g(s)+\rho s\right) \right]  \label{brom1}
\end{equation}
where the contour parallels the imaginary axis with its real part, $s_0$, to
the right of all singularities of the integrand. Since $f(m<0)\equiv 0$, the
integrand is analytic in the right half plane. Therefore, $s_0$ can assume
any non-negative value. Meanwhile, for $f$  given by  (\ref{fm1}), $s=0$
is a branch point singularity. As we shall see, in the subcritical case
there exists a saddle point at positive $s$ and $s_0$ can be chosen to be
this saddle point, whereas in the critical and supercritical cases the
leading contribution is obtained by wrapping the contour around $s=0$.

First we evaluate (\ref{brom1}) in the limit $L\to \infty $ by the
saddle point method, assuming it exists. Let $h(s)\equiv \rho s+\ln g(s)$.
Then the saddle point equation, $h^{\prime }(s_0)=0$, is 
\begin{equation}
\rho =-g^{\prime }(s_0)/g(s_0)=\rho \left( s_0\right)   \label{rhocon}
\end{equation}
leading us to, e.g., 
\begin{equation}
Z(M,L)\simeq \frac{\exp (Lh(s_0))}{\sqrt{2\pi Lh^{\prime \prime }(s_0)}}.
\label{sada}
\end{equation}
If $\rho <\rho _c\equiv \rho \left( 0\right) $, 
then (\ref{rhocon}) has a solution for $s_0>0$, the saddle point
approximation is valid, no condensation occurs.
Substituting (\ref{sada}) in (\ref{pm1}) we get, for $\rho <\rho _c$ and $m\ll
(\rho _c-\rho )L$, 
$p(m)\simeq f(m){\rm e}^{-s_0 m}$,
recovering the GCE upon
identifying the chemical potential $\mu =s_0$.

We now focus  on the behavior as we approach criticality from the {\em subcritical} regime
and  consider (\ref{rhocon}) for small and positive $\rho _c-\rho $, i.e., small $s_0$. Thus, we
just need the small $s$ behavior of $h(s)$. For $f(m)$ in (\ref{fm1}) with
a noninteger $\gamma $, one can expand, quite generally, the Laplace
transform $g(s)$ of (\ref{fm1}) for small $s$, as 
\begin{equation}
g(s)=\sum_{k=0}^{n-1}(-1)^k\frac{\mu _k}{k!}s^k+bs^{\gamma -1}+\ldots
\label{gsexp}
\end{equation}
Here $n=\mbox{int}[\gamma ]$, $\mu _k$ is the $k^{th}$ moment of 
$f(m)$ (which exists for $k<n$).
The second term of (\ref{gsexp}) is the leading singular part and it can be shown
that  $b=A\,\Gamma (1-\gamma )$. 
Note that $\mu _0=1$ for normalized $f$, 
$\mu _1=-g^{\prime }(0)=\rho _c$ and $\Delta \equiv \sqrt{\mu _2-\mu _1^2}$,
which is the width of the distribution $f$, is finite
if $\gamma >3$. 
The role of $\gamma =3$ is now clear. The next-to-leading term is $s^2$ in
one case and $s^{\gamma -1}$ in the other, so that the saddle-point solution
in (\ref{rhocon}) is given by, $s_0 \simeq (\rho _c-\rho )/\Delta ^2$ for 
$\gamma >3$ and $s_0 \simeq  \left[\frac{(\rho _c-\rho )}{b(\gamma -1)}\right]
^{1/(\gamma -2)}$ otherwise. 
Inserting this behavior in the expression 
(\ref{sada}) gives for $\gamma >3$,
$Z(M,L) \sim \exp(-L(\rho_c-\rho)^2/2\Delta^2)$,
pointing to a system with Gaussian distributions and 
normal fluctuations.  In contrast, for $2<\gamma <3$, we will show that
anomalous fluctuations and non-Gaussians appear. 

For the {\em supercritical} regime ($\rho >\rho _c$), there is
no solution to (\ref{rhocon}) on the positive real axis and
more care is needed to find the asymptotic form of $Z(M,L)$.
Our approach is to use the fact that the integral
(\ref{brom1}) will be dominated by $s\simeq 0$.  Thus we can use the 
small $s$ expansion (\ref{gsexp}) and develop  a scaling analysis
by  identifying
the different scaling
regimes and calculating the corresponding scaling forms for
$Z(M,L)$ in the large $L$ limit.

Using 
(\ref{gsexp}),
one can rewrite (\ref{pm1}) as
\begin{equation}
p(m)\simeq f(m)\frac{W\left( (m-M_{ex})/L\right) }{W\left( \rho _c-\rho
\right) },  \label{pm2}
\end{equation}
where 
\[
W(y)=\int_{-i\infty }^{i\infty }\frac{ds}{2\pi i}\exp \left[ L(-ys+\frac{\Delta ^2}{2}s^2+\ldots +b\,s^{\gamma -1})\right] 
\]
and 
$M_{ex}\equiv (\rho -\rho _c)L$ is the excess mass.
All crucial information about the condensate `bump' is
encoded in the asymptotic behavior of $W$. Deferring the details to a later
publication \cite{tbp}, we outline our main results below.
Again, we consider the two cases ($\gamma >$ or $<3$) separately.

Case-I ($\gamma >3$): We find that, for large $L$,
and in the $O(L^{1/2})$ neighborhood of $M_{ex}$, the condensate appears in $p(m)$ as a
pure Gaussian and can be cast in scaling form: 
\begin{equation}
p_{{\rm cond}}(m)\simeq \frac 1{\sqrt{2\pi L^3}\Delta }e^{-z^2/2};\quad
z\equiv \frac{m-M_{ex}}{\Delta L^{1/2}}.  \label{scaling1}
\end{equation}
Note that its integral over $m$ is $1/L$, indicating that the condensation
occurs at a single site.

Case-II ($2<\gamma <3$): 
In the neighborhood of the excess mass 
$M_{ex}$, we find that $p(m)$ has the scaling form:
\begin{equation}
p_{{\rm cond}}(m)\simeq L^{-\gamma /(\gamma -1)}V_\gamma \left[ \frac{%
m-M_{ex}}{L^{1/(\gamma -1)}}\right] .  \label{num1}
\end{equation}
where $V_\gamma (z)=\int_{-i\infty }^{i\infty }\frac{ds}{2\pi i}%
e^{-zs+bs^{\gamma -1}}$. Though we have no closed form for $V_\gamma (z)$,
we obtain its asymptotics: 
\begin{eqnarray}
V_\gamma (z) &\simeq &A\,|z|^{-\gamma }\quad {\rm as}\,\,z\to -\infty 
\label{vzn} \\
&\simeq &c_1\,z^{(3-\gamma )/{2(\gamma -2)}}\,e^{-c_2z^{(\gamma -1)/(\gamma
-2)}}\quad {\rm as}\,\,z\to \infty   \label{vzp}
\end{eqnarray}
where $c_1$, $c_2$ are constants dependent on  $\gamma$.
Note that this condensate `bump' is far from being gaussian:
it has a highly  asymmetric shape, evidenced by (\ref{vzn}, \ref{vzp}).  
The peak occurs at $m=M_{ex}$ and scales as $%
\sim L^{-\gamma /(\gamma -1)}$. Meanwhile its width is $\gamma $ dependent: $%
L^{1/(\gamma -1)}$. The area under the bump is  $1/L$, again
implying that the condensate occurs at only one site.

Finally, in both cases I,II the supercritical partition function is given by
\begin{equation}
Z(M,L)\sim AL/M_{ex}^\gamma
\label{Zsc}
\end{equation}
Similar results for the critical case $\rho =\rho _c$ and $\gamma
=2,3$ (where one obtains logarithmic corrections)
will be published elsewhere \cite{tbp}. 

The implication of these
results are clear: In the condensed phase, the condensate acts as a
reservoir for the critical fluid. Thus, the width of the condensate
bump reflects the mass fluctuation in the critical fluid. For $\gamma
>3$ we showed that condensate is gaussian distributed with width
$\Delta L^{1/2}$, so that the masses in the fluid fluctuate
normally. For $\gamma <3$, however, the width of the condensate and
the mass fluctuations in the fluid are both anomalously large, namely,
$O(L^{1/(\gamma -1)})$. Further implications concern the dynamics
within the steady state. In systems with symmetry breaking, the `flip
time' $\tau $ \cite{EFGM} is of interest. Here the translational 
symmetry is broken by the selection of a site to  hold the condensate
and $\tau $ corresponds to the typical time a condensate exists before
dissolving and reforming on another site. A rough estimate for $\tau $
is $p_{{\rm cond}}^{-1}(m)$, with $|m-M_{ex}|\sim O(L)$. For $\gamma
>3$, this implies flip times growing exponentially with the system size,
whereas for $\gamma <3$, they would diverge more slowly, as some power
of $L$.


Our results may be naturally interpreted
within the framework of sums of
random variables. The partition function (\ref{Zcan}) is proportional to the
probability that the sum of $L$ independent random variables $m_i$, each
distributed according to $f(m)$ is equal to $M$. Given 
$f(m)\sim A m^{-\gamma}$ for large $m$, the $m_i$ are thus L\'{e}vy
flights and $Z(M,L)$ is just proportional to
the probability distribution of the position of
a L\'{e}vy walker (only taking positive steps) after $L$ steps. Then
(\ref{Zsc}) can be interpreted in terms of the extreme
statistics of a L\'{e}vy walk. The mean of a sum of $L$ random variables is
just $\mu _1L$. Thus if $M<L\mu_1 \equiv M_c$ one expects the sum to contain random
variables of typical size $O(1)$, whereas for $M>M_c$ one expects the sum to
be dominated by a rare event i.e. $L-1$ of the variables would be of order
$O(\mu _1)$ except for one which would be large and equal to $M-M_c$. Given
that the distribution of the random variables is $\sim Am^{-\gamma }$, the
probability that this large variable takes the value $M-M_c$ is $%
A\,(M-M_c)^{-\gamma }$. This large contribution could be any of the $L$
possible ones, thus the total probability is $A\,L(M-M_c)^{-\gamma }$,
recovering (\ref{Zsc}).
Moreover, the gaussian $L^{1/2}$ fluctuation for $\gamma >3$ and non-gaussian 
$L^{1/(\gamma -1)}$ fluctuation for $2<\gamma <3$ correspond respectively to
the normal and the anomalous diffusion of a L\'{e}vy process.

To summarise we have considered a very broad class of mass transport
models and derived the condition for condensation. We have presented
an analysis within the canonical ensemble that elucidates the nature
and structure of the condensate. In particular we have identified two
distinct condensate regimes where the condensate is normal and
anomalous and derived the scaling distribution for the two types of
condensate.  Our results rely on the factorisation property of the
steady state (\ref{prodm1}), but we believe the phase scenario of
Fig.2 may apply in models without factorised steady states; it would
be of interest to verify this.  The underlying dynamics of the model
we have studied are one-dimensional however the condition for a
factorised state may be generalized to higher dimensions \cite{tbp}
and work is in progress to investigate condensation in such systems.
Our results confirm that condensation may occur in a wider
class of continuous mass models as well as discrete mass models such
as the ZRP, as suggested in \cite{ZS}. It would also be of 
interest to use this class of models to generalize the phase separation
criterion of  \cite{KLMST} which is  based on the ZRP condensation.

This research is supported in part by the US 
NSF
through DMR-0088451 and DMR-0414122.


\end{multicols}

\end{document}